# Wireless sensors networks MAC protocols analysis

Lamia CHAARI and Lotfi KAMOUN

**Abstract**—Wireless sensors networks performance are strictly related to the medium access mechanism. An effective one, require non-conventional paradigms for protocol design due to several constraints. An adequate equilibrium between communication improvement and data processing capabilities must be accomplished. To achieve low power operation, several MAC protocols already proposed for WSN. The aim of this paper is to survey and to analyze the most energy efficient MAC protocol in order to categorize them and to compare their performances. Furthermore we have implemented some of WSN MAC protocol under OMNET++ with the purpose to evaluate their performances.

——————————— ◆ ———————————

## 1 INTRODUCTION

THE wireless sensor networks (WSNs) are used in a wide range of applications to capture, gather and analyze live environmental data [1]. The wireless sensor network architecture, as described in the majority of the literature [2],[3][4] typically consists of a large number of miniature battery-powered sensor nodes scattered over an area of interest and forming a multi-hop communication network. Each of these nodes has the capacity to collect relevant data from the environment and to transfer them to the sinks nodes that transmits these data then by Internet or by satellite to the central processing node, in order to analyse these data and to take the adequate decisions. In a wireless sensor network sensors nodes are a low cost, resource constrained devices and are often positioned randomly. In many applications they are placed in inaccessible locations, making battery replacement unfeasible. As a consequence, energy efficiency is an important requirement in a medium access control protocol for most wireless sensor networks. Radio energy consumption is a major component contributing to the overall energy consumption at each node.

Many reasons related to MAC paradigms lead to energy waste and WSN life reduction, such as:

- Idle listening: a node doesn't know when will be receiving a frame so it must maintain permanently its radio in the ready to receive mode, as in the DCF method of the wireless networks protocol (IEEE 802.11). This mode consumes a lot of energy, nearly equal to the one consumed in receipt mode.

This energy is wasted if there isn't any transmission on the channel. As in the sensors networks of, the channel is most of the time free. The passive listens presents one of the major reasons of energy loss.

- Collisions: they concern the MAC contention protocols. A collision can occur when a node receives two signals or more simultaneously from different sources that transmit at the same time. When a collision occurs, the energy provided for frame transmission and reception is lost. Let's note that although there are MAC protocols that don't produce any collisions, as TDMA (Time Division Multiple Access), the contention protocols are more used in multi-hops networks because of their simplicity and of their capacity to operate in a decentralized context.

- Overhearing: occurs when a node receives packets that are not destined to him or redundant broadcast.

- Protocol Overhead: can have several origins as the energies lost at the time of transmission and reception of the control frames. For example, the RTS/CTS (Request To Send /Clear To Send) used by some protocols transport no information whereas their transmission consumes energy. Note that the traffic generated by control frames in sensors network is far from being negligible, it could represent until 70% of the global traffic [5].

- Overmitting: occurs when a sensor node sends data to a recipient who is not ready to receive them. Indeed, the sent messages are considered useless and consume an additional energy.

- Packets size: The size of the messages has an effect on the energy consumption of the emitting and receiving nodes. Thus, the size of the packets must not be too elevated nor too weak. Indeed, if it is small, the number of control packets increases the overhead. In the other case, a high transmission power is necessary for large size packets.

————————————————

- *L. Chaari is with Department of Electronic and Information Technology Laboratory (LETI) at Sfax National Engineering School Tunisia.*
- *L. Kamoun is is with Department of Electronic and Information Technology Laboratory (LETI) at Sfax National Engineering School Tunisia.*

- Traffic fluctuation: The fluctuations of the traffic load can lead to the waste a node's energy reserves. Therefore, the protocol should be traffic adaptive.

In [6], the authors introduce a comprehensive node energy model, which includes energy components for radio switching, transmission, reception, listening, and sleeping.

Several researchers have already proposed MAC protocols for wireless sensor networks for achieving energy efficient protocols and their common goal is to reduce sources of the energy waste. In section II of this paper an overview of WSNs energy-efficient MAC protocols is given and taxonomy is elaborated. The rest of the paper is organized as follows. Section III describes the implementation of some WSNs MAC protocol under OMNET++ and shows some experimental results that confirm and validate the previous theoretical results. Finally, conclusions are drawn in section IV.

## 2 ENERGY-EFFICIENCY WSNs MAC PROTOCOLS

MAC protocols for WSNs must guarantee efficient access to the communication media while carefully managing the energy budget allotted to the node. The latter is typically achieved by switching the radio to a low-power mode based on the current transmission schedule. According to channel access policies, most of the existing protocols fall in two categories [7]: contention-based and TDMA-based protocols.

### 2.1 Contention-based MAC

Contention-based MAC protocols are mainly based on the Carrier Sense Multiple Access (CSMA) or Carrier Sense Multiple Access/ Collision Avoidance (CSMA/CA). The main idea is listening before transmitting. The purpose of listening is to detect if the medium is busy, also known as carrier sense. The typical contention-based MAC protocols are S-MAC [8], T-MAC [9], DMAC[17], TEEM[18], UMAC [10] and BMAC[21].

- *Sensor-MAC:* As a slotted energy-efficient MAC protocol, S-MAC [8] is a low-power RTS-CTS protocol for WSNs inspired by 802.11. S-MAC includes four major components: periodic listening and sleeping, collision avoidance, overhearing avoidance, and message passing. After the sleep period, the nodes wake-up and listen whether communication is addressed to them, or they initiate communication themselves. This implies that the sleep and listen periods should be (locally) synchronized between nodes. Each active period is of fixed size, 115 ms, with a variable sleep period. The length of the sleep period dictates the duty cycle of S-MAC. At the beginning of each active period, nodes exchange synchronization information. Following the SYNC period, data may be transferred for the remainder of the active period using RTS-CTS. The advantages of S-MAC are energy waste caused by idle listening is reduced by sleep schedules and time synchronization overhead may be prevented by sleep schedule announcements. Although S-MAC achieves low power operation, it doesn't meet simple implementation, scalability, and tolerance to changing network conditions. As the size of the network increases, S-MAC must maintain an increasing number of neighbors' schedules or incur additional overhead through repeated rounds of resynchronization.

In S-MAC, a node that has more data to send can monopolize the wireless radio channel. This is unfair for other nodes that have short packets to send but need to wait for the completion of the transmission of the long packet.

Many other MAC protocols have been proposed recently which are based on, or inspired by, S-MAC [11][12][13][14][15][16]. S-MAC requires some nodes to follow multiple sleep schedules causing them to wake up more often than the other nodes. In [15], the authors add adaptive listening when a node overhears a neighbor's RTS or CTS packets, it wakes up for a short period of time at the end of their neighbor's transmission to immediately transmit its own data. By changing the duty cycle, S-MAC can trade off energy for latency. In [14] a modification of the S-MAC protocol is proposed (S-MACL) to eliminate the need for some nodes to stay awake longer than the other nodes. The modified version improves the energy efficiency and increases the life span of a wireless sensor network. In [16], an adaptive S-MAC was proposed that enables low duty cycle operation, common sleep schedules to reduce control overhead, low latency and traffic adaptive wakeup. To reduce control overhead and latency, it introduces coordinated sleeping among neighboring nodes.

In [22], the author's compare adaptive-rate MACs to adaptive power schemes.

- *Time out -MAC:* As declared above, the SMAC protocol does not work well when the traffic load fluctuates. To overcome this problem, the TMAC protocol [9] introduces the timeout value to finish the active period of a node. If a node does not hear anything within the period corresponding to the time-out value, it allows the node to go into sleep state.

T-MAC, in variable workloads, uses one fifth the power of S-MAC. In homogeneous workloads, TMAC and S-MAC perform equally well. T-MAC suffers from the same complexity and scaling problems of S-MAC. Shortening the active window in T-MAC reduces the ability to snoop on surrounding traffic and adapt to changing network condition.

- *DMAC[17]:* The DMAC could be summarized as an improved Slotted Aloha algorithm in which slots are assigned to the sets of nodes based on a data gathering tree. During the receive period of a node, all of its child nodes have transmit periods and contend for the medium. It can achieve very good latency compared to other sleep/listen period. However, collision avoidance methods are not utilized in DMAC. Hence, when a number of nodes that have the same schedule try to send to the same node, collisions will occur.

- *UMAC:* It is based on the SMAC protocol and provides three main improvements on this protocol, e.g. various duty-cycles, utilization based tuning of duty-cycle, selective sleeping after transmission. The scheme does not as-sign the same duty cycle for nodes, and each node can be assigned different periodically listen and sleep schedules with different duty cycle. Utilization based tuning of duty-cycle reflects to different traffic loads of every node in a network. Such variation corresponds to the diversity of performed tasks by a particular node and its location. Selective sleeping after transmission avoids the above energy wastage. A node should go to sleep "selectively". When transmission is finished, a node checks if it is at scheduled sleep time, and goes to sleep if it's at scheduled sleep time. It does not introduce additional delays, since traffic is not expected to this node. In consequence, the proposed protocol improves energy efficiency as well as end-to-end latency.

*-Traffic Aware Energy Efficient MAC (TEEM) [18]:* TEEM makes two important modifications over the existing S-MAC protocol: firstly by having all nodes turn off their radios much earlier when no data packet transfer is expected to occur in the networks, and secondly by eliminating communication of a separate RTS control packet even when data traffic is likely to occur. The listen period In TEEM consists of Syncdata and Syncnodata. The first part of the listen period in TEEM contains data while the other part contains no data. Both parts are used for synchronization. Each node will listen in the first Syncdata part of its listen period whether someone has data to transfer or not. If there is no data in the Syncdata part then it will send its own sync packet in the Syncnodata part. The TEEM protocol combines the Sync and RTS packets into one packet called SyncRTS. Whenever a node wants to communicate with another node, it sends the SyncRTS packet in its Syncdata part. The destination node receives the packet and starts the communication, while the other nodes synchronize themselves with a SyncRTS packet and go into sleep mode. TEEM MAC is a good choice in small networks because there are fewer chances of retransmission.

*- Berkeley Media Access Control( B-MAC)[21]*: B-MAC uses clear channel assessment (CCA) and packet backoffs for channel arbitration, link layer acknowledgments for reliability, and low power listening (LPL). B-MAC makes local policy decisions to optimize power consumption, latency, throughput, fairness or reliability. To achieve low power operation, BMAC employs an adaptive preamble sampling scheme to reduce duty cycle and minimize idle listening (an adaptive rate scheme). B-MAC supports on-the-fly reconfiguration and provides bidirectional interfaces for system services to optimize performance, whether it is for throughput, latency, or power conservation. By comparing B-MAC to S-MAC, we see that B-MAC's flexibility results in better packet delivery rates, throughput, latency, and energy consumption than S-MAC.

## 2.2 TDMA-based MAC

Although random access achieves good flexibility and low latency for applications with low traffic loads, deterministic scheduling is actually the most effective way of eliminating the sources of energy waste. With perfect scheduling, only one transmitter-receiver pair would be active during each transmission period, therefore, reducing collision and eliminating idle-listening and overhearing. Use of TDMA is viewed as a natural choice for sensor networks because radios can be turned off during idle times in order to conserve energy However, deterministic TDMA scheduling1 requires a large overhead in order to maintain accurate synchronization between sensors and to exchange local information, such as the network topology and the communication pattern. Furthermore, the latency increases linearly with the total number of sensors sharing the channel since TDMA assigns a separate time-slot to each transmitting sensor.

*- EYES MAC[19]:* The TDMA-based EMACs protocol divides time into time slots, which nodes can use to transfer data without having to content for the medium or having to deal with energy wasting collisions of transmissions. A node can assign only one slot to itself and is said to control this slot. After the frame length, which consists of several time slots, the node again has a period of time reserved for it.

A time slot is further divided in three sections: Communication Request (CR), Traffic Control (TC) and the data section. In the CR section other nodes can do requests to the node that is controlling the current time slot. Nodes that have a request will pick a random start time in the short CR section to make their request. The controller of a time slot will always transmit a TC message in the time slot. When a time slot is not controlled by any node, all nodes will remain in sleep state during that time slot. The time slot controller also indicates in its TC message what communication will take place in the data section. If a node is not addressed in the TC section nor its request was approved, then the node will resume in standby state during the entire data section. The TC
message can also indicate that the controlling node is about to send an omnicast message. After the TC section the actual data transfer takes place.

*- Lightweight MAC [20] :* This protocol is based on ideas of the EMACs. LMAC protocol takes into account the physical layer properties. The intension of the protocol is to minimize the number of transceiver switches, to make the sleep interval for sensor nodes adaptive to the amount of data traffic.

During its time slot, a node will always transmit a message which consists of two parts: control message and a data unit.

The control message has a fixed size and is used for several purposes. It carries the ID of the time slot controller, it indicates the distance of the node to the gateway in hops for simple routing to a gateway in the network, it addresses the intended receiver and reports the length of the data unit.

The control data will also be used to maintain synchronization between the nodes and therefore the nodes also transmit the sequence number of their time slot in the frame. The transmission of the control data is carefully timed by the nodes, although we do not assume that the nodes have clocks with high accuracy. All neighboring nodes will put effort in receiving the control messages of their neighboring nodes. When a node is not addressed in that message or the message is not addressed as an omnicast message, the nodes will switch off their power consuming transceivers only to wake at the next time slot. If a node is addressed, it will listen to the data unit which might not fill the entire remainder of the time slot. Both transmitter and receiver(s) turn off their transceivers after the message transfer has completed.

- *Advanced Medium Access Control (A-MAC) [25]:* is a TDMA-based MAC protocol developed for low rate and reliable data transportation with the view of prolonging the network lifetime, adapted from LMAC protocol. Compared to conventional TDMA-based protocols, which depend on central node manager to allocate the time slot for nodes within the cluster, the AMAC protocol uses distributed technique where node selects its own time slot by collecting its neighborhood information. The protocol uses the supplied energy efficiently by applying a scheduled power down mode when there is no data transmission activity.

The protocol is structured into several frames, where each frame consists of several time slots. Each node transmits a beacon message at the beginning of its time slot, which is used for two purposes; as synchronization signal and neighbor information exchanges. By using this message, the controlled node informs which of its neighboring nodes will be participating in the next data session. The intended nodes need to stay in listening mode in order to be able to receive the intended packet, while other nodes turn to power down mode until the end of current time slot. The time slot assignment in A-MAC is divided into four states; initial, wait, discover, and active. A new node that enters a network starts its operation in initial state where node listens to the channel for its neighbor's beacon message in order to synchronize with the network. Node starts synchronization when it receives a beacon message from one of its neighbors and adjusts its timer by subtracting the beacon received time with beacon transmission time. Node remains in this state for aListenFrame frames in order to find the strongest beacon signal. Before entering the wait state, node randomly chooses a number of waiting frame. Node enters the discover state when the waiting counter expired and start collecting its neighborhood information by listening for its neighboring node's beacon messages for a period of aListen-Frame frames. Node enters active state when it successfully selects a time slot. Node enters sleep mode in two scenarios. First, after transmitting a beacon message and no more data packet scheduled to be transmitted. Second, if received beacon message from it neighboring node indicates no incoming data packet. Compared to LMAC, A-MAC allows node to transmit to multiple destinations.

TDMA requires strict synchronization among users and a centralized control to coordinate the use of the channels. Benefitting from the extra coordination, it is easier for TDMA to achieve the users' QoS demands, e.g. the rate, delay or bit-error-rate (BER) requirements, while consuming less resources. Even with the complexity of computing the optimal channel allocation and the increase of control messages, it is often worth-while for delay-constrained or energy constrained applications. In addition, the coordination also allows TDMA to achieve better throughput under heavy traffic loads.

Although energy waste on collisions has been avoided, there are a number of negative aspects. Cluster, which is used in TDMA-based MAC, is difficult to dynamically change its frame length and time slot assignments, thus contributes to poor scalability. TDMA-based protocols require strict time synchronization which results in high cost on hardware and high latency for data.

Recently there are others research that there have been some hybrid proposals (ZMAC[23], GMAC[24]), which combine the advantages of contention-based MAC with that of TDMA-based MAC. However hybrid MAC protocols are usually complex in transition mechanisms between contention-based and TDMA-based, in addition, these protocols are more complex in implementation.

## 3 WSNs MAC PROTOCOL PERFORMANCE ANALYSIS

In this section we have used discrete event simulator OMNet++ we implemented LMAC and BMAC with a framework for wireless networks.

### 3.1 LMAC performance analysis

LMAC, in state of sleep, the node always accepts packets of the network layer which permits to adapt the node sleep interval with the data traffic. We consider the network as shown in Fig. 1. Table 1 illustrates simulation parameters.

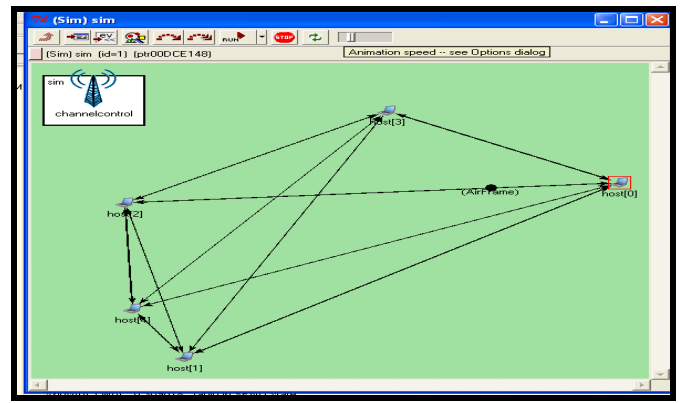

Fig. 1. A sample LMAC network topology

TABLE I
LMAC simulation parameters

| Control Channel | |
|---|---|
| carrierFrequency[en Hz] | 868e+6 |
| Pmax [mw] | 1.0 |
| Seuil du signal d'atténuation [dBm] | -110 |
| Coefficient alpha | 3.0 |
| **LMAC Layer** | |
| HeaderLength [en bit] | 24 |
| queuLength [en bit] | 50 |
| numSlots | 4 |
| Bit rate [bit/seconde] | 19200 |

| slotDuration [s] | 0.2 |
|---|---|
| controlDuration [s] | 0.02 |

The "EndToEndDelayVec" which represents message lifetime at the host is shown in fig;2.

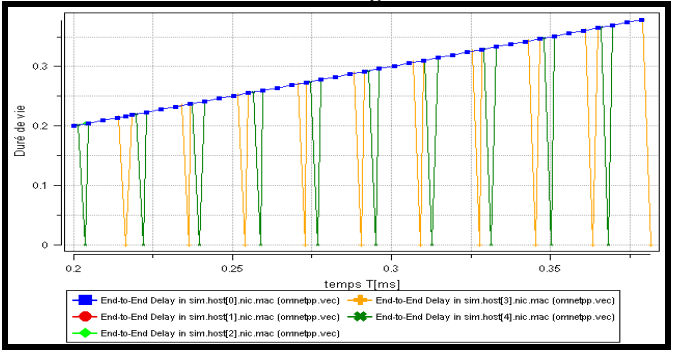

Fig. 2. LMAC lifetime

## 3.2 BMAC performance analysis

BMAC node can be in one of the following three states: packets transmission state, channel verification state and sleep state. We study the effect of BMAC parameters (slot duration and bit rate) on the messages life time. Simulation results are shown respectively in fig.3 and fig.4.

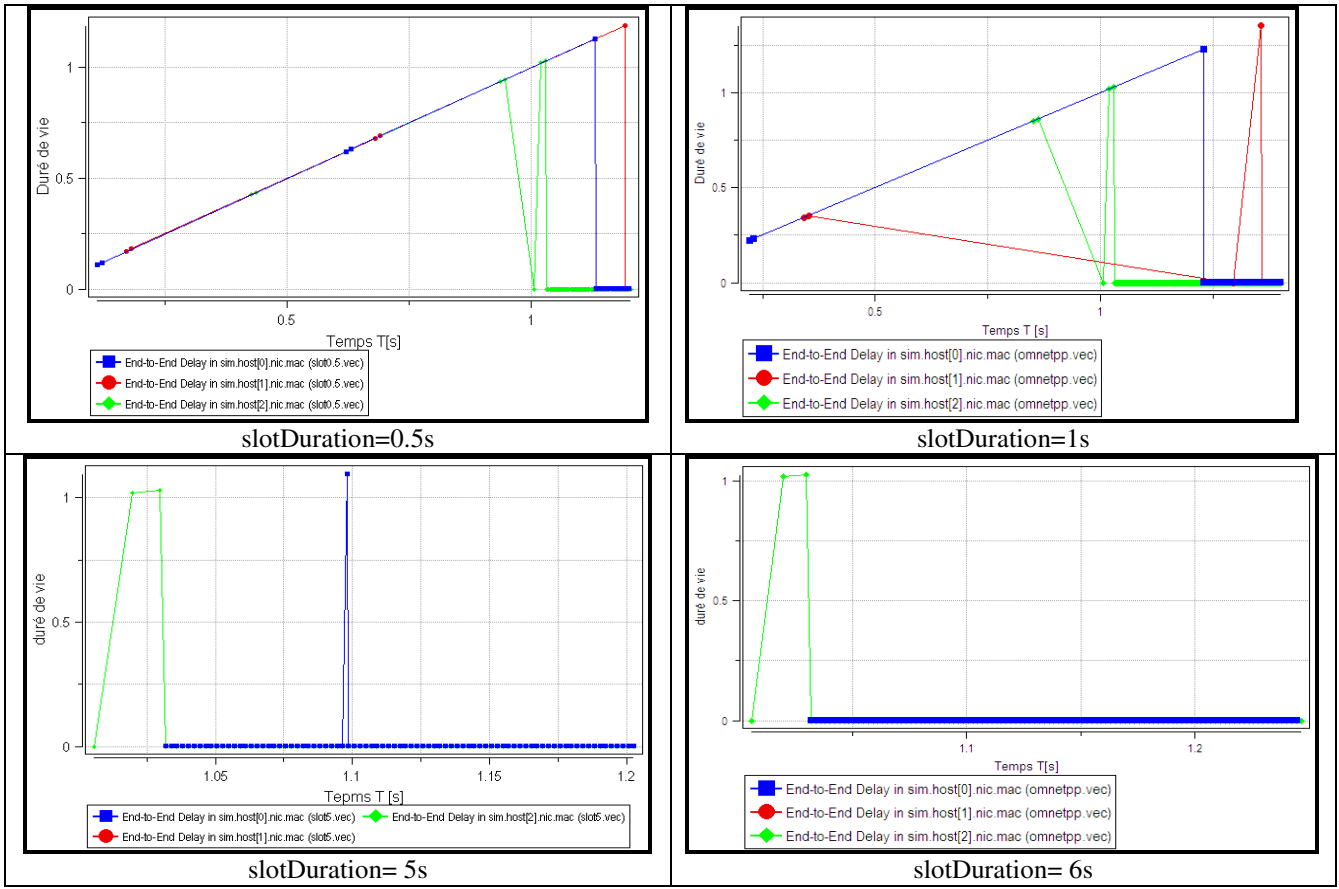

Fig. 3. LMAC lifetime (slot duration)

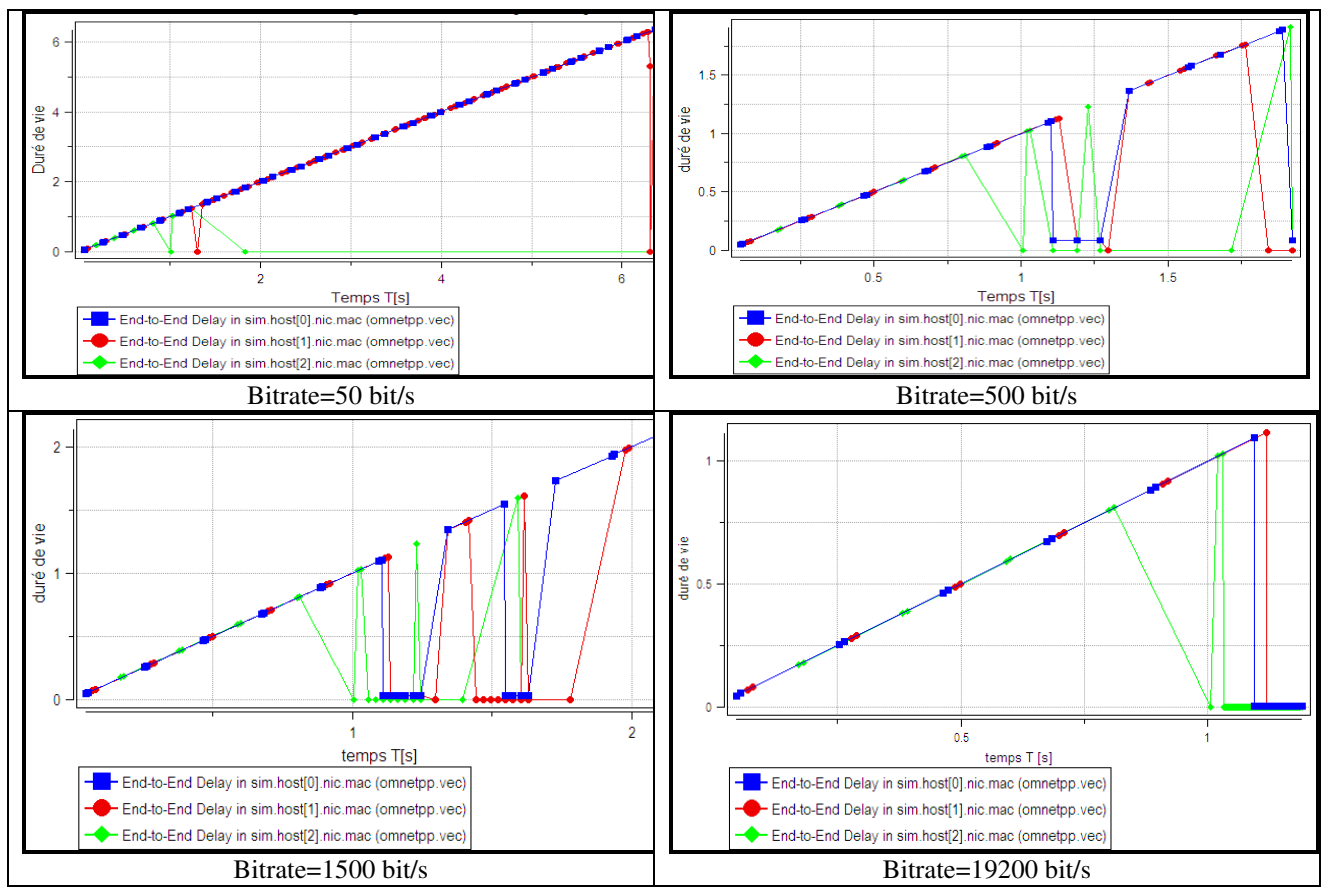

Fig. 4. LMAC lifetime (Birate)

## CONCLUSION

This paper formulates the MAC problem, in the context of minimizing energy utilization, in wireless sensor communications. Two prominent MAC protocols used for many applications, random access and Time Division Multiple Access (TDMA), were studied. Next, our study was validated via the implementation in OMNET++ of some MAC protocol.

Our survey of WSN MAC protocols and their concepts shows that, despite many proposals, no perfect proposal has been published. However adaptive and cross layer designs solutions can leads to achieve both high performance and low energy consumption at the same time.

The design of optimal WSN MAC protocol with optimal parameters must take as input the application specifications (network topology and packet generation rate), the application requirements for energy consumption, delay and reliability, and the constraints from the physical layer (energy consumption and data rate), so cross layer design for WSN MAC protocol could be an interesting issue.

**First A**. Author    Dr Lamia CHAARI was born in Sfax, Tunisia, in 1972. She received the engineering and PhD degrees in electrical and electronic engineering from Sfax national engineering school (ENIS) in TUNISIA. Actually she is an assistant professor in multimedia and informatics higher institute in SFAX She is also a researcher in electronic and technology information laboratory (LETI). Her scope of research are communications, networking and signal processing which are specially related to wireless and new generation networks.

**Second B. Author Jr.** Lotfi Kamoun was born in Sfax Tunisia, 25 January. 1957. He received the electrical engineering degree from the Sciences and Techniques Faculty in Tunisia. Actually he is a Professor in Sfax national engineering school (ENIS) in TUNISIA. He is the director of   electronic and technology information laboratory (LETI). His scope of research are communications, networking, Software radio and signal processing which are  specially related to wireless and new generation networks.